\documentclass[conference]{IEEEtran}
\usepackage{amsmath, amssymb, bm, cite, epsfig, psfrag}
\usepackage{graphicx}
\usepackage{soul} 
\usepackage[margin=0.95in]{geometry}
\usepackage{dblfloatfix}
\usepackage{array}
\usepackage{lipsum}

\newcolumntype{P}[1]{>{\centering\hspace{0pt}}p{#1}}
\newcolumntype{M}[1]{>{\centering\hspace{0pt}}m{#1}}
\newcolumntype{L}{>{\centering\arraybackslash}m{3cm}}
\usepackage[font=small]{caption}
\usepackage{epstopdf}	
\usepackage{longtable}
\usepackage{supertabular,booktabs}
\usepackage{bbm}
\usepackage{multirow}
\usepackage[usenames,dvipsnames]{xcolor}

\usepackage{subfigure}
\usepackage{etoolbox}
\usepackage{pbox}
\usepackage{fixltx2e}
\usepackage{tabu}	
\usepackage{filecontents}
\usepackage{enumerate}
\usepackage{textcomp}
\usepackage{makecell}
\usepackage{gensymb}
\usepackage{colortbl}
\usepackage{fancyhdr}

\usepackage{bm}

\newtoggle{conference}

\togglefalse{conference} 
\graphicspath{{figures/}}

\newcolumntype{?}{!{\vrule width 2pt}}

\fancypagestyle{firststyle}{
	\fancyhf{}
	\fancyhead[L]{S. Ju, Y. Xing, O. Kanhere and T. S. Rappaport, ``3-D Statistical Indoor Channel Model for Millimeter-Wave and Sub-Terahertz Bands,''  2020 \textit{IEEE Global Communications Conference (GLOBECOM)}, Dec. 2020, pp. 1-7.}     

}
\IEEEoverridecommandlockouts

\begin{document}
\title{3-D Statistical Indoor Channel Model for Millimeter-Wave and Sub-Terahertz Bands}
\author{\IEEEauthorblockN{Shihao Ju, Yunchou Xing, Ojas Kanhere, and Theodore S. Rappaport}

\IEEEauthorblockA{	\small NYU WIRELESS, NYU Tandon School of Engineering, Brooklyn, NY, 11201\\
				\{shao, ychou, ojask, tsr\}@nyu.edu}
}

\maketitle
\thispagestyle{firststyle}

\begin{abstract} 

Millimeter-wave (mmWave) and Terahertz (THz) will be used in the sixth-generation (6G) wireless systems, especially for indoor scenarios. This paper presents an indoor three-dimensional (3-D) statistical channel model for mmWave and sub-THz frequencies, which is developed from extensive channel propagation measurements conducted in an office building at 28 GHz and 140 GHz in 2014 and 2019. Over 15,000 power delay profiles (PDPs) were recorded to study channel statistics such as the number of time clusters, cluster delays, and cluster powers. All the parameters required in the channel generation procedure are derived from empirical measurement data for 28 GHz and 140 GHz line-of-sight (LOS) and non-line-of-sight (NLOS) scenarios. The channel model is validated by showing that the simulated root mean square (RMS) delay spread and RMS angular spread yield good agreements with measured values. An indoor channel simulation software is built upon the popular NYUSIM outdoor channel simulator, which can generate realistic channel impulse response, PDP, and power angular spectrum.

\end{abstract}
    

\section{Introduction}
Driven by ubiquitous usage of mobile devices and the emergence of massive Internet of Things (IoT), the sixth-generation (6G) wireless system will offer unprecedented high data rate and system throughput. Moving to millimeter-wave (mmWave) and Terahertz (THz) frequencies (i.e., 30 GHz - 3 THz) is considered as a promising solution of fulfilling future data traffic demand created by arising wireless applications such as augmented/virtual reality (AR/VR) and 8K ultra high definition (UHD) due to the vast bandwidth \cite{Rap19access}. 

Accurate THz channel characterization for indoor scenarios facilitates the designs of transceivers, air interface, and protocols for 6G and beyond. Standards documents like IEEE 802.11 ad/ay and 3GPP TR 38.901 proposed statistical channel models for various indoor scenarios such as office, home, shopping mall, and factory \cite{80211ad10,80211ay16,3GPP38901r16}. The IEEE 802.11 ad/ay developed a double directional statistical channel impulse response (CIR) model working at 60 GHz based on measurements and ray-tracing results in the conference room, cubical environment, and living room \cite{80211ad10,80211ay16}. 3GPP TR 38.901 presented a cluster-based statistical channel model for frequencies from 0.5 to 100 GHz for outdoor and indoor scenarios with different values of large-scale parameters such as delay spread, angular spread, Rician K-factor, and shadow fading \cite{3GPP38901r16}. \textcolor{black}{This paper proposes a unified indoor channel model across mmWave and sub-THz bands based on indoor channel measurements at 28 GHz and 140 GHz, which provides a reference for future standards development above 100 GHz.}

There has been some work on channel modeling at THz bands \cite{Priebe13twc, Han2015twc,Zajic20access}. A temporal-spatial statistical channel model based on ray-tracing results in an office room at 275-325 GHz, which consisted of line-of-sight (LOS), first- and second-order reflected paths, was developed in \cite{Priebe13twc}. A generic multi-ray channel model for 0.06-1 THz constituted by LOS, reflected, diffracted, and scattered paths generated by ray-tracing simulations was given in \cite{Han2015twc}. Due to the hardware constraints such as maximum transmit power, most of the published channel propagation measurements at THz frequencies were limited within a few meters \cite{Zajic20access}. The existing channel models for THz frequencies are mainly ray-tracing based, where CIRs are represented as a superposition of LOS, reflected, and scattered paths based on reflection and scattering properties of the environment. \textcolor{black}{In this paper, the presented three-dimensional (3-D) channel model for mmWave and sub-THz frequencies is statistical and built upon extensive propagation measurements up to 45 m in an office building at 28 GHz and 140 GHz.}

The remainder of the paper is organized as follows. Section \ref{sec:desp} describes the indoor measurements at 28 GHz and 140 GHz performed in 2014 and 2019. Section \ref{sec:pl} introduces the large-scale path loss model. Section \ref{sec:method} describes the spatial statistical CIR model, and Section \ref{sec:stat} shows statistics of required parameters in the channel generation procedure. Section \ref{sec:simulation} validates the proposed channel model by showing that simulated and measured omnidirectional root mean square (RMS) delay spread (DS) and global RMS angular spread (AS) yield a good agreement. Section \ref{sec:conclusion} provides concluding remarks.

\section{Wideband Channel Measurements} \label{sec:desp}
\subsection{Measurement environment, system, and procedure}
\begin{figure}[h!]
	\centering
	\includegraphics[width=.5\textwidth]{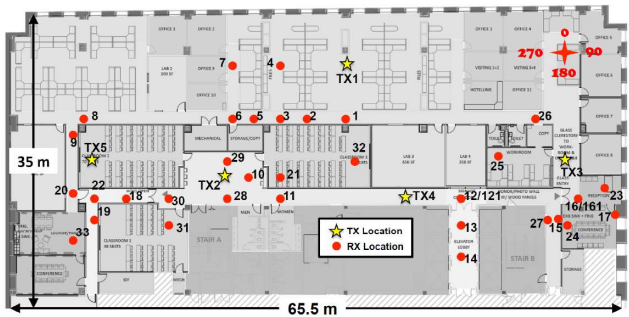}
	\caption{Floor plan of the 9th floor, 2 MetroTech Center. There are five TX locations denoted as yellow stars and 33 RX locations denoted as red dots \cite{Mac15access}. }
	\label{fig:floor_plan}
\end{figure}

28 GHz and 140 GHz channel measurement campaigns were conducted in the NYU WIRELESS research center on the 9th floor of 2 MetroTech Center in downtown Brooklyn, New York in 2014 and 2019, respectively. The 9th floor is a typical office environment (65.5 m $\times$ 35 m $\times$ 2.7 m) with common obstructions such as cubical partitions, corridors, conference rooms, walls made of drywall, glass doors, and desks. Five transmitter (TX) locations and 33 receiver (RX) locations were chosen for 28 GHz measurements resulting in 10 LOS TX-RX location pairs and 38 NLOS TX-RX location pairs as seen in Fig. \ref{fig:floor_plan}. The measured TX-RX separation distance for 28 GHz measurements ranged from 3.9 m to 45.9 m. Three TX locations (TX1, TX2, and TX5) were chosen for 140 GHz measurements, and only 16 out of the 33 RX locations were measured due to the limited transmit power (0 dBm), resulting in five LOS TX-RX location pairs and 12 NLOS TX-RX location pairs. The measured TX-RX separation distance for 140 GHz measurements ranged from 3.9 m to 39.2 m. 

We used a wideband sliding correlation-based channel sounding system with highly directional horn antennas to record PDPs using a high-speed oscilloscope \cite{Mac17sounder}. The channel sounder transmitted pseudorandom sequence signals with 800 MHz and 1 GHz RF bandwidth at 28 GHz and 140 GHz, respectively. 
Two identical horn antennas were used at TX and RX, which had 15 dBi gain and 30\degree~antenna half power beamwidth (HPBW) for 28 GHz measurements, and 27 dBi gain and 8\degree~antenna HPBW for 140 GHz measurements. For each TX-RX location pair, eight unique antenna azimuth sweeps were performed to obtain spatial statistics for vertical-to-vertical (V-V) polarization. Overall, at most 96 (= $8 \times 12$) at 28 GHz and 360 (= $8 \times 45$) at 140 GHz directional PDPs were acquired with V-V polarization configuration for each TX-RX location pair.

For each antenna azimuth sweep, the RX (TX) horn antenna was swept in steps of 8\degree~or 30\degree~according to antenna HPBW, and the pointing direction of TX (RX) horn antenna was fixed during one sweep. At each sweeping step, an average PDP over 20 instantaneous PDPs was recorded by a high-speed oscilloscope for next-step processing and analysis. By sweeping antennas in the azimuth plane at different elevation levels, multipath components (MPCs) with energy above the noise floor that can arrive at RX in the 3-D space were captured and recorded. More details of the measurement procedure can be found in \cite{Mac15access,Xing20jsac}.
\subsection{Data processing}
Omnidirectional channel modeling is fundamental and regarded as a basis for directional and multiple input multiple output (MIMO) channel modeling \cite{Samimi16mtt}. Thus, omnidirectional PDPs need to be recovered from measured directional PDPs with knowledge of absolute time delays. Note that the channel sounding system was not able to perform precise synchronization between TX and RX, thus cannot provide absolute timing information because the PDP recording was triggered at the first MPC arrival and only had excess time delay information. Here we used a ray-tracing tool to provide the actual time of flight of the first MPC in measured directional PDPs so that an omnidirectional PDP can be synthesized by aligning directional PDPs in time at each TX-RX location pair. 

A 3-D mmWave ray-tracing tool, NYURay \cite{Kanhere19globecom}, was used to predict possible propagating rays for the identical TX-RX location pairs selected in measurements, which can provide time of flight of the predicted rays. If the direction of a measured directional PDP matched the direction of a predicted ray, the time of flight of this predicted ray was used as the absolute time delay of the first MPC in this measured PDP. Due to the beamwidth of horn antennas, there might be several predicted rays which were close to the pointing angle of the measured PDP in space and vice versa. The predicted ray which was closest to the measured PDP in 3-D space (i.e., the minimal sum of the differences of azimuth angle of departure (AOD) $\phi_{\textup{AOD}}$, zenith angle of departure (ZOD) $\theta_{\textup{ZOD}}$, azimuth angle of arrival (AOA) $\phi_{\textup{AOA}}$, and zenith angle of arrival (ZOA) $\theta_{\textup{ZOA}}$) was chosen to match this measured PDP and provide the absolute timing information. 

\section{\textcolor{black}{Large-scale Path Loss Model}}\label{sec:pl}
\textcolor{black}{Path loss models describe the distance-power law that the received power decreases exponentially with distance and are commonly used in the prediction of signal strength and cell range. A popular path loss model, close-in free space reference distance (CI) path loss model with 1 m reference distance, is given by \cite{Sun16tvt,Mac15access, Rap15tcomm}
	\begin{equation}
	\label{eq:pathloss}
	\begin{split}
	\textup{PL}^{\textup{CI}}(f,d)[\textup{dB}]=&\textup{FSPL}(f,1\:\textup{m})[\textup{dB}]+10n\log_{10}(d)+\chi_\sigma,
	\end{split}
	\end{equation}
	where $n$ is path loss exponent (PLE) and $d$ is the distance. $\textup{FSPL}(f,1\:\textup{m})$ is the free space path loss at 1 m at frequency $f$. Shadow fading $\chi_\sigma$ is characterized by a zero mean Gaussian random variable with standard deviation $\sigma$ in dB.}

\begin{figure}[h!]
	\centering
	\includegraphics[width=1\linewidth]{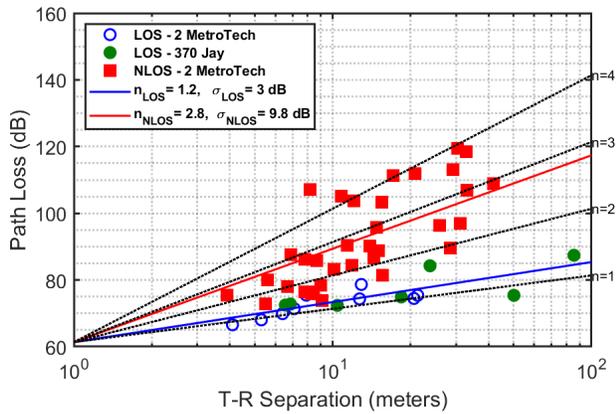}
	\caption{\textcolor{black}{28 GHz indoor omnidirectional path loss scatter plot and MMSE-fitted CI path loss model with distance for LOS and NLOS scenarios.}}
	\label{fig:pl28}
\end{figure}

\textcolor{black}{PLE $n$ indicates that the power decays by 10$n$ dB per decade of distance beyond 1 m \cite{Rap15tcomm}. Fig. \ref{fig:pl28} shows that PLE for 28 GHz LOS and NLOS scenarios via minimum mean square error (MMSE) fitting are 1.2 and 2.8, respectively. 1.2 is derived from 2 MetroTech dataset for 28 GHz LOS case, which is lower than 1.3-1.9 found in the literature \cite{Ko16a,Xu02jsac,Rubio19a}. To verify this low PLE, we conducted LOS measurements in another office building, 370 Jay, at 28 GHz. The resulting PLE for 370 Jay dataset is also 1.2, suggesting that the power attenuates much slower (12 dB per ten meters) than values reported in the literature, which might be attributed to the strong waveguide effect of long and narrow corridors in the indoor environment.}

\section{Statistical Omnidirectional Channel Impulse Response Model} \label{sec:method}
\textcolor{black}{Here we presented a 3GPP-like spatial statistical channel model for indoor scenarios using the NYUSIM outdoor channel modeling framework \cite{Samimi16mtt}. The outdoor and indoor NYUSIM channel model share the channel generation procedure and the set of channel parameters, but have different probabilistic distributions for these channel parameters such as the number of time clusters (TCs) and cluster subpaths (SPs) due to the distinct environment characteristics. The proposed indoor channel model is built upon the existing NYUSIM outdoor modeling procedure and developed from 28 GHz and 140 GHz measurements in the office building.}
	
\textcolor{black}{The NYUSIM and 3GPP TR 38.901 channel models are both 3-D statistical channel models but have several key differences \cite{3GPP38901r16, Sun18tvt}.} The 3GPP channel model defines a cluster as a group of MPCs closely spaced jointly in the temporal and spatial domain \cite{3GPP38901r16} while the NYUSIM channel model separately characterizes temporal and spatial statistics by defining time cluster and spatial lobe (SL) motivated by the observations from measurements \cite{Mac15access}. A TC is composed of MPCs traveling close in time, and that arrive from potentially different directions in a short propagation time window. An SL represents a main direction of arrival or departure where MPCs may arrive over hundreds of ns \cite{Samimi16mtt}. 

Both modeling methodologies are valid, where the 3GPP model is more widely used and the NYUSIM model has a simpler and physically-based structure \cite{Polese17jsac,Barati15twc}. Performance evaluation with respect to spectrum efficiency, coverage, and hardware/signal processing requirements between the 3GPP and NYUSIM channel models were analyzed in detail \cite{Sun18tvt}. 

A received signal can be regarded as a superposition of multiple replicas of the transmitted signal with different delays and angles for any wireless propagation channel. The time cluster spatial lobe (TCSL)-based omnidirectional CIR model is given by 
\begin{equation}
\label{eq:omni_cir}
\begin{split}
h_{\text{omni}}(t,\Theta,\Phi) = &\sum_{n=1}^{N}\sum_{m=1}^{M_n}a_{m,n}e^{j\varphi_{m,n}}\cdot\delta(t-\tau_{m,n})\\
&\cdot\delta(\overrightarrow{\Theta}-\overrightarrow{\Theta_{m,n}}) \cdot\delta(\overrightarrow{\Phi}-\overrightarrow{\Phi_{m,n}}),
\end{split}
\end{equation}
where $t$ is the absolute propagation time, $\overrightarrow{\Theta}=(\phi_{\textup{AOD}},\theta_{\textup{ZOD}})$ is the vector of AOD and ZOD, and $\overrightarrow{\Phi}=(\phi_{\textup{AOA}},\theta_{\textup{ZOA}})$ is the vector of AOA and ZOA. $N$ and $M_n$ denote the number of TCs and the number of cluster SPs, respectively. For the $m$th SP in the $n$th TC, $a_{m,n}$, $\varphi_{m,n}$, $\tau_{m,n}$, $\overrightarrow{\Theta_{m,n}}$, and $\overrightarrow{\Phi_{m,n}}$ represent the magnitude, phase, absolute propagation time, AOD and AOA, respectively. The PDP and power angular spectrum (PAS) can be obtained by integrating the square of the CIR in space and time domains, respectively. 

The measured PDP and PAS were partitioned into time clusters and spatial lobes to obtain empirical channel statistics such as the number of TCs and the number of SLs, respectively \cite{Samimi16mtt}. The partition in the temporal domain was realized by minimum inter-cluster time void interval (MTI). Two sequentially recorded SPs belong to two distinct time clusters if the difference of the excess time delays of these two SPs is beyond MTI. For example, 25 ns was used as MTI for an outdoor urban microcell (UMi) environment \cite{Samimi16mtt}, while 6 ns was used as MTI in this paper for an indoor office environment due to the fact that the width of a typical hallway in the measured indoor office environment was about 1.8 m (i.e., $\sim$6 ns propagation delay). 

The partition in the spatial domain was realized by spatial lobe threshold (SLT). The angular resolution of the measured PAS depends on the HPBW of horn antennas. To reconstruct the 3-D spatial distribution of the received power, linear interpolation in azimuth and elevation planes with 1\degree~resolution was implemented using the measured directional received powers. A power segment was calculated every 1\degree~direction in the 3-D space using this linear interpolation. Neighboring power segments above the SLT formed an SL. A -10 dB SLT below the maximum peak power in the 3-D power spectrum was applied, which was the same as in \cite{Samimi16mtt}.

\section{Statistics Extraction of Channel Parameters} \label{sec:stat}
Temporal and spatial channel parameters are extracted from the measured PDP and PAS using the TCSL approach described in Section \ref{sec:method}. Temporal parameters include the number of TCs ($N$) and SPs in a TC ($M_n$), TC excess delay ($\tau_n$) and intra-cluster SP excess delay ($\rho_{m,n}$), TC power ($P_n$) and SP power ($\Pi_{m,n}$). Spatial parameters are the number of SLs ($L$), the mean azimuth and elevation angle of an SL ($\phi$ and $\theta$) and the azimuth and elevation angular offset of a SP ($\Delta \phi$ and $\Delta \theta$) with respect to the mean angle of the SL. Values of parameters required in Table \ref{tab:model} for 28 GHz and 140 GHz LOS and NLOS scenarios are given in Table \ref{tab:values}, where DU stands for discrete uniform.

\begin{table*}[]
	\centering
	\caption{\textsc{Input Parameters for channel coefficient generation procedure}}
	\label{tab:model}
	\begin{tabular}{|c|c|c|c|}
		\hline
		\textbf{Step Index} & \textbf{Channel Parameters} & \multicolumn{2}{c|}{\textbf{28 - 140  GHz InH}} 
		\\ \hline
		\textbf{Step 1}  & \textbf{\# Time clusters $N$}  & \multicolumn{2}{c|}{$N \sim  \begin{cases} \textup{DU}(1,N_c) &,\textup{if LOS}\\ \textup{Poisson}(\lambda_c)&,\textup{if NLOS}\end{cases}$}  \\ \hline
		\textbf{Step 2}  &\textbf{\# Cluster subpaths $M_n$}  & \multicolumn{2}{c|}{$M_n \sim (1-\beta)\delta(M_n)+\beta \cdot \textup{DE}(\mu_s)$}
		\\ \hline
		\textbf{Step 3} &\textbf{Intra-cluster delay $\rho_{m,n}$ (ns)}   & \multicolumn{2}{c|}{$\rho_{m,n}\sim $ Exp$(\mu_{\rho})$}    
		\\ \hline
		\textbf{Step 4}   & \textbf{Cluster delay $\tau_n$ (ns)}                                  & \multicolumn{2}{c|}{$\!\begin{aligned}\tau_n' &\sim \textup{Exp}(\mu_{\tau}) \:\text{or}\: \textup{Logn}(\mu_{\tau},\sigma_{\tau})\\ \Delta \tau_n&=\textup{sort}(\tau_n')-\textup{min}(\tau_n')\\ \tau_n &= \begin{cases} 0 & ,n=1\\ \tau_{n-1}+\rho_{M_{n-1},n-1}+\Delta\tau_n+\textup{MTI} &,n=2,...N \end{cases}\end{aligned}$}                                                                                                                                                                                                                                                                             \\ \hline
		\textbf{Step 5}   & \textbf{Cluster power $P_n$ (mW)}   & \multicolumn{2}{c|}{$\!\begin{aligned}	P'_n &= \bar{P}_0 e^{-\frac{\tau_n}{\Gamma}}10^{\frac{Z_n}{10}}, \label{eq:cp1}\\
			P_n &= \frac{P'_n}{\sum_{k=1}^N P'_k} \times P_r [mW], \label{eq:cp2}\\
			Z_n&\sim \mathcal{N}(0, \sigma_Z), \;n = 1,2,...,N \label{eq:cp3}\end{aligned}$} \\ \hline
		\textbf{Step 6}  & \textbf{Subpath power $\Pi_{m,n}$(mW)}      & \multicolumn{2}{c|}{$\!\begin{aligned}
			\Pi'_{m,n} &= \bar{\Pi}_0 e^{-\frac{\rho_{m,n}}{\gamma}}10^{\frac{U_{m,n}}{10}}, \label{eq:sp1}\\
			\Pi_{m,n} &= \frac{\Pi'_{m,n}}{\sum_{k=1}^{M_n} \Pi'_{k,n}} \times P_n [mW], \label{eq:sp2}\\
			U_{m,n}&\sim \mathcal{N}(0, \sigma_U), \;m = 1,2,...,M_n \label{eq:sp3}
			\end{aligned}$}                                                                                                                                                                                                                             \\ \hline
		\textbf{Step 7}   & \textbf{SP phase $\varphi$ (rad)}      &\multicolumn{2}{c|}{Uniform(0, 2$\pi$)}        \\ \hline
		\textbf{Step 8}   & \textbf{\# Spatial lobes $L$}   & \multicolumn{2}{c|}{\begin{tabular}[c]{@{}c@{}} $L_{\textup{AOD}}\sim \text{DU}(1,L_{\textup{AOD,max}})$\\$L_{\textup{AOA}}\sim \text{DU}(1,L_{\textup{AOA,max}})$
		\end{tabular}}          \\ \hline
		\textbf{Step 9} & \textbf{SL mean angle $\phi_i,\theta_i$ (\degree)}    &\multicolumn{2}{c|}{\begin{tabular}[c]{@{}c@{}} $\phi_i\sim \textup{Uniform}(\phi_{\textup{min}},\phi_{\textup{max}}),$\\$\phi_{\textup{min}}=\frac{360(i-1)}{L},\phi_{\textup{max}}=\frac{360i}{L},i=1,2,...,L$\\$\theta_i\sim \mathcal{N}(\mu_{l},\sigma_{l})$
		\end{tabular}}                                                                                                                                          \\ \hline
		\textbf{Step 10} & \makecell{\textbf{SP angle offset $\Delta\phi_i,\Delta\theta_i$}\\ \textbf{ w.r.t $\phi_i,\theta_i$ (\degree)}} &\multicolumn{2}{c|}{$\!\begin{aligned}
			i\sim\textup{DU}[1,L_{\textup{AOD}}]&,j\sim\textup{DU}[1,L_{\textup{AOA}}]\\
			(\Delta\phi_i)_{m,n,\textup{AOD}}&\sim \mathcal{N}(0,\sigma_{\phi,\textup{AOD}})\\
			(\Delta\theta_i)_{m,n,\textup{ZOD}}&\sim \mathcal{N}(0,\sigma_{\theta,\textup{ZOD}}) \\
			(\Delta\phi_j)_{m,n,\textup{AOA}}&\sim \mathcal{N}(0,\sigma_{\phi,\textup{AOA}})\\
			(\Delta\theta_j)_{m,n,\textup{ZOA}}&\sim \mathcal{N}(0,\sigma_{\theta,\textup{ZOA}})\\
			\end{aligned}$}                                                                                                 \\ \hline
	\end{tabular}
\end{table*}

\subsection{Temporal Channel Parameters}
\subsubsection{The number of time clusters}
\begin{figure}
	\centering
	\includegraphics[width=.95\linewidth]{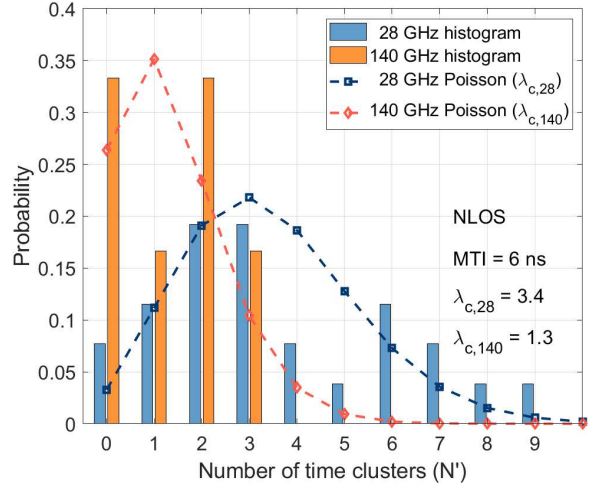}
	\caption{The number of time clusters for 28 GHz and 140 GHz NLOS scenario follows Poisson distributions with mean 3.4 and 1.3, respectively.}
	\label{fig:num_tc_nlos_28_140}
\end{figure}

The empirical histogram of the number of TCs $N$ from 28 GHz and 140 GHz NLOS measurements with a 6 ns MTI is shown in Fig. \ref{fig:num_tc_nlos_28_140}, which follows a Poisson distribution. Since Poisson distribution starts from zero while the number of TCs is at least one, $N'=N-1$ is used for distribution fitting. The generated number of TCs from the Poisson distribution is added by one to obtain the simulated number of TCs, which is given by
\begin{equation}
\begin{split}
P(N'=k)& = \frac{\lambda_c^k}{k!}e^{-\lambda_c},\quad\quad k=0,1,2,...,\\
N &= N'+1.
\end{split}
\end{equation}
where $\lambda_c$ is the mean of the Poisson distribution. Fig. \ref{fig:num_tc_nlos_28_140} shows that there are more time clusters at 28 GHz than 140 GHz which is likely due to the higher partition loss at 140 GHz (e.g., 4-8 dB higher than 28 GHz) \cite{Xing20jsac}. The Poisson distribution of the number of TCs for the indoor NLOS scenario is different from the uniform distribution for the outdoor scenario \cite{Samimi16mtt}. 

\subsubsection{Number of cluster subpaths}
\begin{figure}
	\centering
	\includegraphics[width=.95\linewidth]{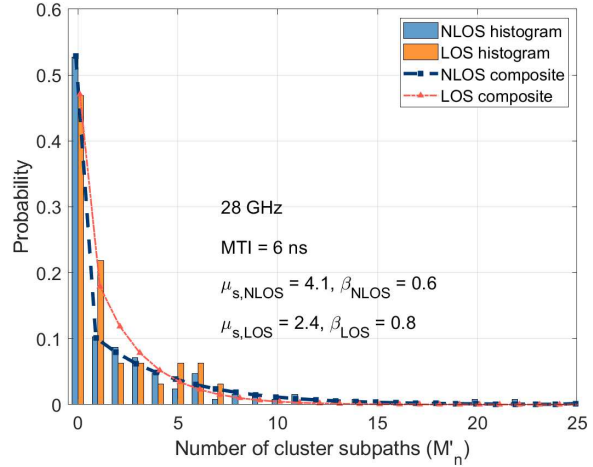}
	\caption{The number of cluster subpaths for 28 GHz LOS and NLOS scenarios follows the composite distribution. }
	\label{fig:num_sp_nlos_28}
\end{figure}
The number of cluster SPs $M_n$ is related to the number of TCs and depends on the choice of MTI. \textcolor{black}{A larger MTI causes fewer TCs and more SPs per TC since two SPs might be counted into one cluster rather than two clusters using a larger MTI.} The empirical histogram of the number of cluster subpaths from 28 GHz measurements with a 6 ns MTI is shown in Fig. \ref{fig:num_sp_nlos_28}. Similar to the number of time clusters, $M_n'=M_n-1$ was used for distribution fitting. We proposed a composite distribution with a $\delta$-function at $M_n'=0$ and a discrete exponential (DE) distribution, which is given by
\begin{equation}
\begin{split}
P_{M_n'}(k) = (1-\beta)\delta(k)&+\beta\int_{k}^{k+1}\frac{1}{\mu_s}e^{-\frac{x}{\mu_s}}dx,\\
&k=0,1,2,...,
\end{split}
\end{equation}
where $\mu_s$ is the mean of the DE distribution and $\beta$ is the weight of the DE distribution in the composite distribution. The maximum likelihood estimation (MLE) of $\mu_s$ and $\beta$ are 4.1 and 0.6 for 28 GHz NLOS measurements, respectively. The identical composite distribution for the 28 GHz LOS scenario shows that $\mu_s=2.4$ and $\beta=0.8$, suggesting that NLOS scenario forms relatively larger clusters than the LOS scenario. \textcolor{black}{As shown in Table \ref{tab:values}, in terms of the number of TCs (depending on the input parameter $\lambda_c$) and the number of SPs (depending on the input parameter $\mu_s$), the 140 GHz channel is much sparser than the 28 GHz channel, which is critical for channel estimation \cite{Huang19tsp}.}

\subsubsection{Inter- and intra-cluster excess delay}
The cluster excess delay $\tau_n$ is defined as the time difference between the first arriving subpath in the PDP and the first arriving subpath in the $n$th cluster, as shown in Step 4 of Table \ref{tab:model}. $\rho_{M_{n-1},n-1}$ is the intra-cluster excess delay of the last subpath in the former cluster. $\Delta \tau_n$ is the inter-cluster excess delay without MTI (i.e., 6 ns). Results given in Table \ref{tab:values} show that the inter-cluster delays for 28 GHz NLOS, 140 GHz LOS, and 140 GHz NLOS datasets are well characterized by exponential distributions. However, the inter-cluster delay for 28 GHz LOS dataset is closer to a lognormal distribution. The peculiar behavior found in 28 GHz LOS scenario is attributed to two strong MPCs with large inter-cluster delays observed in the corridor environment (e.g., TX4 and RX 16).

The intra-cluster excess delay $\rho_{m,n}$ is defined as the time difference between the first arriving subpath and the $m$th arriving subpath in the $n$th time cluster. An exponential distribution with $\mu_{\rho}$ shows a good agreement with the measured intra-cluster excess delay for 28 GHz and 140 GHz LOS and NLOS scenarios.

\subsection{Spatial Channel Parameters}
\subsubsection{The number of spatial lobes} An SL represents a main direction of arrival or departure. The angular resolution of the measured PAS depends on the antenna HPBW (30\degree~and 8\degree~for 28 and 140 GHz measurements). Linear interpolation of the measured PAS with 1\degree~angular resolution in the azimuth and elevation planes was implemented to investigate the 3-D spatial distribution of the received power. Measurement results showed that there are at most two main directions of arrival in the azimuth plane, except that there are a few NLOS locations having three main directions of arrival in 28 GHz measurements. 

\subsubsection{Spatial lobe mean angle and subpath angular offset}
The SL mean azimuth angle $\phi_i$ is generated by first dividing the azimuth plane into a few sectors according to the number of spatial lobes, then randomly selecting a direction within that sector. The SL mean elevation angle $\theta_i$ is generated by a normal random variable with the mean $\mu_l$. The angular offsets of each subpath in both azimuth and elevation planes are modeled as zero-mean normal random variables with variances $\sigma_{\phi}$ and $\sigma_{\theta}$.   
\begin{table*}[]
	\centering
	\caption{\textsc{Required parameters that reproduce the measured statistics of omnidirectional channels using the presented statistical channel model.}}
	\label{tab:values}
	\begin{tabular}{|c|c|c|c|c|}
		\hline
		\textbf{Input Parameters}                                                          & \textbf{28 GHz LOS}   & \textbf{28 GHz NLOS}  & \textbf{140 GHz LOS}  & \textbf{140 GHz NLOS} \\ \hline
		$N_c$                                                                   & 5                     & NA                     & 4                    & NA                     \\ \hline
		$\lambda_c$                                                               & NA                     & 3.4                   & NA                     & 1.3                   \\ \hline
		$\beta_s$                                                                 & 0.8                   & 0.6                   & 0.8                   & 1.0                   \\ \hline
		$\mu_s$                                                                   & 2.4                   & 4.1                   & 1.0                   & 1.0                   \\ \hline
		$\mu_{\tau}[\textup{ns}]$                                                 & logn(2.7, 1.4)        & 12.1                  & 18.6                  & 23.5                   \\ \hline
		$\mu_{\rho}[\textup{ns}]$                                                 & 2.6                   & 15.7                  & 2.2                   & 2.2                    \\ \hline
		$\Gamma[\textup{ns}],\sigma_{Z}[\textup{dB}]$                            & 38.7, 5.0             & 20.1, 7.0             & 6.0, 3.0             & 13.4, 5.0              \\ \hline
		$\gamma[\textup{ns}],\sigma_{U}[\textup{dB}]$                             & 2.5, 7.0              & 5.0, 8.0            &    1.4, 5.0           &     2.0, 6.0           \\ \hline
		$L_{\textup{AOD,max}},L_{\textup{AOA,max}}$                                                        & 2, 2                     &  2, 3                    &       2, 2              &  2, 2              \\ \hline
		$\mu_{l,\textup{ZOD}}[\degree],\sigma_{l,\textup{ZOD}}[\degree]$                 & -7.3, 3.8                   & -5.5, 2.9                   & -6.8,  4.9                   & -2.5, 2.7                    \\ \hline
		$\mu_{l,\textup{ZOA}}[\degree],\sigma_{l,\textup{ZOA}}[\degree]$                 & 7.4, 3.8                   & 5.5, 2.9                   & 7.4, 4.5                   & 4.8, 2.8                    \\ \hline
		$\sigma_{\phi,\textup{AOD}}[\degree],\sigma_{\theta,\textup{AOD}}[\degree]$ & 23.5, 16.0          & 31.6, 15.6            & 4.8, 4.2              & 5.1, 4.1                \\ \hline
		$\sigma_{\phi,\textup{AOA}}[\degree],\sigma_{\theta,\textup{AOA}}[\degree]$ & 19.3, 14.5          & 25.5, 14.6            & 4.8, 4.3             & 5.4, 4.2                \\ \hline
	\end{tabular}
\end{table*}

\section{Simulation Results} \label{sec:simulation}
The statistical channel model presented in Section \ref{sec:stat} was implemented as an indoor channel simulator based on the NYUSIM outdoor channel simulator to investigate the accuracy of the simulated temporal and spatial statistics by comparing it with the measured statistics. Note that the parameters listed in Table \ref{tab:values} are primary statistics which are used in the channel parameter generation procedure. The metrics used in this section for channel validation are secondary statistics such as RMS DS and RMS AS which are not explicitly used in the channel generation, but the simulated secondary statistics should yield good agreements with the measured statistics. 10,000 simulations were carried out for four frequency scenarios (i.e., 28 GHz LOS, 28 GHz NLOS, 140 GHz LOS, and 140 GHz NLOS) by generating 10,000 omnidirectional PDPs, and 3-D AOD and AOA PASs as sample functions of (\ref{eq:omni_cir}) using the NYUSIM indoor channel simulator.

\subsection{Simulated RMS Delay Spreads}
\begin{figure}[h!]
	\centering
	\includegraphics[width=1\linewidth]{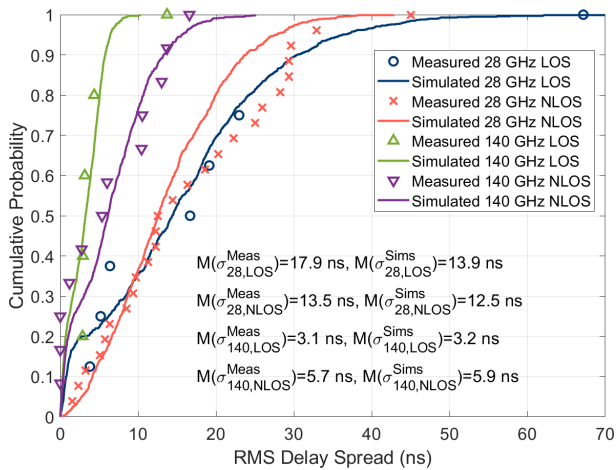}
	\caption{Measured and simulated omnidirectional RMS DS for 28 and 140 GHz in LOS and NLOS scenarios. }
	\label{fig:rms_ds_4cases}
\end{figure}

The RMS DS describes channel temporal dispersion, which is a critical metric to validate a statistical channel model. Fig. \ref{fig:rms_ds_4cases} shows the simulated and measured omnidirectional RMS DS at 28 GHz and 140 GHz in LOS and NLOS scenarios. As shown in Fig. \ref{fig:rms_ds_4cases}, the empirical and simulated median RMS DS are 17.9 and 13.9 ns for 28 GHz LOS scenario, 13.5 and 12.5 ns for 28 GHz NLOS scenario, 3.1 and 3.2 ns for 140 GHz LOS scenario, and 5.7 and 5.9 ns for 140 GHz NLOS scenario, respectively. The simulated cumulative distribution function (CDF) yielded good agreements with the empirical CDF for four frequency scenarios. 

\subsection{Simulated RMS Angular Spreads}
\begin{figure}[]
	\centering
	\includegraphics[width=1\linewidth]{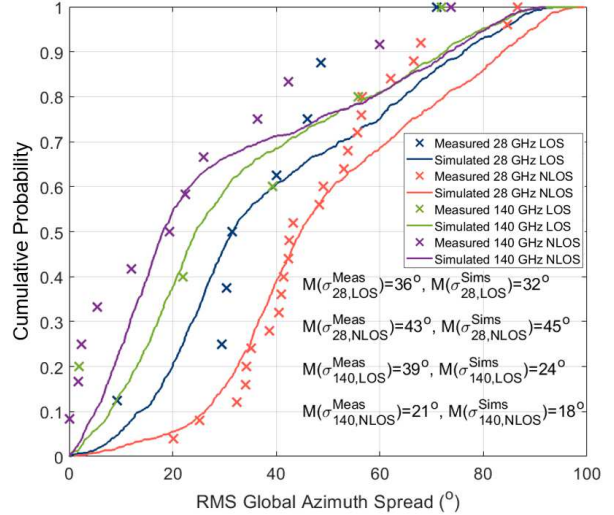}
	\caption{Measured and simulated global RMS AOA AS for 28 and 140 GHz in LOS and NLOS scenarios. }
	\label{fig:rms_azi_as_nlos}
\end{figure}

The omnidirectional azimuth and elevation AS describe the angular dispersion at a TX or RX over the entire 4$\pi$ steradian sphere, also termed global AS. The AOA and AOD global ASs were computed using the total (integrated over delay) received power over all measured azimuth/elevation pointing angles. The measured and simulated global AOA RMS AS was calculated using Appendix A-1,2 in \cite{3GPP38901r16} and compared in Fig. \ref{fig:rms_azi_as_nlos}, showing the simulated and measured median global angular spreads are very close (less than 5\degree) for 28 GHz LOS, 28 GHz NLOS, and 140 GHz NLOS cases while the limited number of 140 GHz LOS measurements considerably skewed the empirical distribution.  

\section{Conclusion} \label{sec:conclusion}
The paper presented a 3-D spatial statistical channel model based on the extensive measurements at 28 and 140 GHz in an indoor office building. The statistics of necessary parameters for the channel generation procedure were extracted from empirical LOS and NLOS measurement data. The NYUSIM indoor channel simulator was used to generate thousands of PDPs and PASs for validating the presented channel model. The simulated channel statistics yielded a good agreement with the measured channel statistics in terms of secondary statistics such as omnidirectional RMS DS and global RMS AS. The omnidirectional statistical channel model in this work will be used as a basis for further directional and MIMO channel modeling and facilitates channel estimation and channel capacity analysis for mmWave and sub-THz frequencies.

\bibliographystyle{IEEEtran}
\bibliography{twc}

\begin{thebibliography}{10}
\providecommand{\url}[1]{#1}
\csname url@samestyle\endcsname
\providecommand{\newblock}{\relax}
\providecommand{\bibinfo}[2]{#2}
\providecommand{\BIBentrySTDinterwordspacing}{\spaceskip=0pt\relax}
\providecommand{\BIBentryALTinterwordstretchfactor}{4}
\providecommand{\BIBentryALTinterwordspacing}{\spaceskip=\fontdimen2\font plus
\BIBentryALTinterwordstretchfactor\fontdimen3\font minus
  \fontdimen4\font\relax}
\providecommand{\BIBforeignlanguage}[2]{{%
\expandafter\ifx\csname l@#1\endcsname\relax
\typeout{** WARNING: IEEEtran.bst: No hyphenation pattern has been}%
\typeout{** loaded for the language `#1'. Using the pattern for}%
\typeout{** the default language instead.}%
\else
\language=\csname l@#1\endcsname
\fi
#2}}
\providecommand{\BIBdecl}{\relax}
\BIBdecl

\bibitem{Rap19access}
T.~S. {Rappaport} \emph{et~al.}, ``Wireless communications and applications
  above 100 {GHz}: Opportunities and challenges for {6G} and beyond,''
  \emph{IEEE Access}, vol.~7, pp. 78\,729--78\,757, June 2019.

\bibitem{80211ad10}
A.~Maltsev \emph{et~al.}, ``Channel models for 60 {GHz WLAN} systems,'' doc.:
  IEEE 802.11-09/0334r8, May 2010.

\bibitem{80211ay16}
------, ``Channel models for {IEEE} 802.11ay,'' doc.: IEEE 802.11-15/1150r9,
  May 2016.

\bibitem{3GPP38901r16}
3GPP, ``Technical specification group radio access network; study on channel
  model for frequencies from 0.5 to 100 {GHz (Release 16)},'' TR 38.901
  V16.0.0, October 2019.

\bibitem{Priebe13twc}
S.~{Priebe} and T.~{Kurner}, ``Stochastic modeling of {THz} indoor radio
  channels,'' \emph{IEEE Transactions on Wireless Communications}, vol.~12,
  no.~9, pp. 4445--4455, September 2013.

\bibitem{Han2015twc}
C.~{Han}, A.~O. {Bicen}, and I.~F. {Akyildiz}, ``Multi-ray channel modeling and
  wideband characterization for wireless communications in the {Terahertz}
  band,'' \emph{IEEE Transactions on Wireless Communications}, vol.~14, no.~5,
  pp. 2402--2412, May 2015.

\bibitem{Zajic20access}
C.~{Cheng}, S.~{Sangodoyin}, and A.~{Zajić}, ``{THz} cluster-based modeling
  and propagation characterization in a data center environment,'' \emph{IEEE
  Access}, vol.~8, pp. 56\,544--56\,558, March 2020.

\bibitem{Mac15access}
G.~R. {Maccartney}, T.~S. {Rappaport}, S.~{Sun}, and S.~{Deng}, ``Indoor office
  wideband millimeter-wave propagation measurements and channel models at 28
  and 73{ GHz} for ultra-dense {5G} wireless networks,'' \emph{IEEE Access},
  vol.~3, pp. 2388--2424, October 2015.

\bibitem{Mac17sounder}
G.~R. {MacCartney} and T.~S. {Rappaport}, ``A flexible millimeter-wave channel
  sounder with absolute timing,'' \emph{IEEE Journal on Selected Areas in
  Communications}, vol.~35, no.~6, pp. 1402--1418, March 2017.

\bibitem{Xing20jsac}
Y.~Xing, T.~S. Rappaport, and A.~Ghosh, ``Millimeter wave and sub-{Terahertz}
  indoor office building propagation measurements and channel models at 28, 73,
  and 142 {GHz},'' \emph{submitted to IEEE Journal on Selected Areas in
  Communications}, April 2020.

\bibitem{Samimi16mtt}
M.~K. {Samimi} and T.~S. {Rappaport}, ``{3-D} millimeter-wave statistical
  channel model for {5G} wireless system design,'' \emph{IEEE Transactions on
  Microwave Theory and Techniques}, vol.~64, no.~7, pp. 2207--2225, July 2016.

\bibitem{Kanhere19globecom}
O.~{Kanhere} \emph{et~al.}, ``Map-assisted millimeter wave localization for
  accurate position location,'' in \emph{2019 IEEE Global Communications
  Conference (GLOBECOM)}, December 2019, pp. 1--6.

\bibitem{Sun16tvt}
S.~{Sun} \emph{et~al.}, ``Investigation of prediction accuracy, sensitivity,
  and parameter stability of large-scale propagation path loss models for {5G}
  wireless communications,'' \emph{IEEE Transactions on Vehicular Technology},
  vol.~65, no.~5, pp. 2843--2860, May 2016.

\bibitem{Rap15tcomm}
T.~S. Rappaport \emph{et~al.}, ``Wideband millimeter-wave propagation
  measurements and channel models for future wireless communication system
  design ({Invited Paper}),'' \emph{IEEE Transactions on Communications},
  vol.~63, no.~9, pp. 3029--3056, September 2015.

\bibitem{Ko16a}
J.~Ko, S.~U. Lee, Y.~S. Kim, and D.-J. Park, ``Measurements and analyses of 28
  {GHz} indoor channel propagation based on a synchronized channel sounder
  using directional antennas,'' \emph{Journal of Electromagnetic Waves and
  Applications}, vol.~30, no.~15, pp. 2039--2054, June 2016.

\bibitem{Xu02jsac}
{Hao Xu}, V.~{Kukshya}, and T.~S. {Rappaport}, ``Spatial and temporal
  characteristics of {60-GHz} indoor channels,'' \emph{IEEE Journal on Selected
  Areas in Communications}, vol.~20, no.~3, pp. 620--630, April 2002.

\bibitem{Rubio19a}
L.~Rubio \emph{et~al.}, ``Wideband propagation channel measurements in an
  indoor office environment at 26 {GHz},'' in \emph{2019 IEEE International
  Symposium on Antennas and Propagation and USNC-URSI Radio Science Meeting},
  October 2019, pp. 2075--2076.

\bibitem{Sun18tvt}
S.~{Sun}, T.~S. {Rappaport}, M.~{Shafi}, P.~{Tang}, J.~{Zhang}, and P.~J.
  {Smith}, ``Propagation models and performance evaluation for {5G}
  millimeter-wave bands,'' \emph{IEEE Transactions on Vehicular Technology},
  vol.~67, no.~9, pp. 8422--8439, June 2018.

\bibitem{Polese17jsac}
M.~{Polese}, M.~{Giordani}, M.~{Mezzavilla}, S.~{Rangan}, and M.~{Zorzi},
  ``Improved handover through dual connectivity in {5G} mmwave mobile
  networks,'' \emph{IEEE Journal on Selected Areas in Communications}, vol.~35,
  no.~9, pp. 2069--2084, June 2017.

\bibitem{Barati15twc}
C.~N. {Barati}, S.~A. {Hosseini}, S.~{Rangan}, P.~{Liu}, T.~{Korakis}, S.~S.
  {Panwar}, and T.~S. {Rappaport}, ``Directional cell discovery in millimeter
  wave cellular networks,'' \emph{IEEE Transactions on Wireless
  Communications}, vol.~14, no.~12, pp. 6664--6678, July 2015.

\bibitem{Huang19tsp}
C.~{Huang}, L.~{Liu}, C.~{Yuen}, and S.~{Sun}, ``Iterative channel estimation
  using lse and sparse message passing for mmwave mimo systems,'' \emph{IEEE
  Transactions on Signal Processing}, vol.~67, no.~1, pp. 245--259, January
  2019.

\end{thebibliography}

\end{document}